# Sodium Ion Ordering in $Na_xCoO_2$

H.W. Zandbergen[1,2], M.L. Foo[1], Q. Xu[2], V. Kumar[2] and R. J. Cava[1]

[1] Department of Chemistry and Princeton Materials Institute, Princeton University, Princeton, NJ 08544 USA

[2] National Centre for HREM, Department of Nanoscience, Delft University of Technology, Rotterdamseweg 137, 2628 AL Delft, The Netherlands

**Abstract**

The layered sodium cobalt oxide $Na_xCoO_2$ is studied by electron diffraction for a wide range of sodium contents, $0.15<x<0.75$. An extensive series of ordered Na ion-Na vacancy superlattices is found beyond the simple hexagonal average structure. The most strongly developed superlattice is found for the composition $Na_{0.5}CoO_2$, which displays $Co^{3+}/Co^{4+}$ charge ordering at low temperatures. The structural principle for some of the observed ordering schemes, particularly near $x=0.5$, is, surprisingly, the presence of lines of Na ions and vacancies rather than simply maximized Na-Na separations.

**1. Introduction**

$Na_xCoO_2$, with a crystal structure consisting of hexagonal layers of Na ions sandwiched between planes of edge shared $CoO_6$ octahedra, has been the subject of extensive study in the past 20 years due to the high mobility of the Na ions. With both high ion mobility and good electronic conductivity, it has been considered as a cathode material in high energy density, reversible solid state battery systems [1-3].

Recent interest has focused on the characterization and understanding of the surprisingly high thermoelectric power of the metallic conductor $Na_{0.7}CoO_2$ [4,5], a composition in the range of the thermodynamically stable sodium stoichiometries for $Na_xCoO_2$, $0.65<x<0.75$. The transport and magnetic properties of $Na_xCoO_2$ are strongly dependent on the number of charge carriers introduced by deviations from stoichiometry in the Na sublattice, and the overall range of Na contents possible in $Na_xCoO_2$ has been reported to be $0.3<x<0.75$, e.g. [6], with the lower Na contents obtained by room temperature deintercalation of sodium from $Na_{0.7}CoO_2$. It has been shown that $Na_xCoO_2$ exhibits metallic behavior for a wide range of both high and low x, but that for $x=1/2$ an insulating state exists[7]. Foo [7] et al have suggested that this insulating behavior is based on strong coupling between the electronic charge carriers (holes) and the Na ions. Interest has also recently increased in $Na_xCoO_2$ due to the discovery of superconductivity for the x=0.35 composition when it is intercalated with water [6], leading to the composition $Na_{0.35}CoO_2 \cdot 1.3H_2O$.

Several structure determinations of $Na_xCoO_2$ have been reported [8-11]. All reported structures are based the stacking of triangular O-Co-O layers with Na planes. The Co atoms are in edge-sharing $CoO_6$ octahedra. The $Na^+$ ions have been shown to have both trigonal prismatic coordination and octahedral coordination with oxygen [1] between the O-Co-O layers. Prismatic coordination occurs when the close packed oxygen planes directly adjacent to the Na plane have the same projection into the basal plane (A-Na-A), whereas octahedral coordination of Na occurs when the directly adjacent oxygen planes have different projections (A-Na-B) into the basal plane. The various reported stacking sequences of the oxygen atoms are given in Table I. In almost all cases, only the average structure has been reported. Very recently Shi et al [17] have reported a 2d[110] superstructure for $Na_xCoO_2$ with 0.75<X<1 due to ordering of Na and Na-vacancies.

Apart from the influence of x on the $Co^{3+}/Co^{4+}$ ratio (i.e. for the hypothetical composition x=0 all $Co^{4+}$ would be found, and for hypothetical x=1, all $Co^{3+}$ would be found) and through that on the physical properties, the presence and type of ordering of the Na ions over the available sites is important for the physical properties of the material, since at certain compositions the lock-in of stable sodium ordered superstructures is expected to influence the properties. As interest in the properties of $Na_xCoO_2$ increases, it is important to characterize possible Na ion ordering and the resulting structural modulation as precisely as possible. In this paper, we present electron diffraction proof of the existence of sodium ordered superstructures over a wide range of sodium contents in $Na_xCoO_2$, and present structural models for that ordering.

**2. Experimental**

$Na_{0.75}CoO_2$ was made by solid state synthesis. Stoichiometric amounts of $Na_2CO_3$ and $Co_3O_4$ corresponding to $Na_{0.77}CoO_2$ were mixed and heated to 800 degrees under flowing $O_2$. Due to volatility of Na, the final Na composition is slightly lower than the starting composition. The sample of $Na_{0.35}CoO_2$ and $Na_{0.30}CoO_2$ were prepared by stirring $Na_{0.75}CoO_2$ (0.5 g) in 20 ml of a 5x or 40x $Br_2$ solution in dry acetonitrile at room temperature for 4 days.

('1x' indicates that the amount of Br$_2$ used is exactly the amount that would theoretically be needed to remove all of the sodium from Na$_{0.7}$CoO$_2$.) The product was washed copiously with acetonitrile and dried under dry Ar. To prepare Na$_{0.5}$CoO$_2$, Na$_{0.75}$CoO$_2$ (0.4 g) was stirred in 60 ml of a 10x solution of I$_2$ in acetonitrile for 4 days. The product was then washed copiously with acetonitrile and then air-dried. Na$_{0.64}$CoO$_2$ was prepared by stirring Na$_{0.75}$CoO$_2$ (0.5 g) in 20 ml of a 0.5x solution of I$_2$ in acetonitrile for 4 days. The product was then washed copiously with acetonitrile and then air-dried. Na$_{0.15}$CoO$_2$ was prepared by reacting Na$_{0.75}$CoO$_2$ (0.15 g) with 1 molar equivalent of NO$_2$PF$_6$ in dry acetonitrile under Ar atmosphere for 2 days with magnetic stirring. The product was then washed with dry acetonitrile and dried under Ar atmosphere. The sodium contents of all samples were determined by multiple independent composition measurements by the ICP-AES method, and are considered to be reliable to +-0.02/formula unit precision.

Electron microscopy was performed on all samples described above. Electron transparent areas were mainly obtained by very gently crushing the specimen under hexane, depositing a few droplets on a holey carbon Cu grid, and insertion of the specimen in the microscope within 1 minute to reduce the possibility of degradation by exposure to air. Also, conventional ion milling and focused ion beam preparation were applied, but too many artifacts were obtained with these methods. Electron diffraction experiments were performed with Philips CM300UT and CM200ST electron microscopes, both equipped with a field emission gun and EDX element analysis equipment. Nanodiffraction was performed using a condenser aperture of 10 μm and an electron probe size of 10-30 nm in diameter. The specimen cooling experiments were performed using Gatan He cooling holder for operation at about 20 K and a custom-made holder for operation at about 100 K. Approximate temperatures are given because in such an apparatus the temperature is not measured on the crystal that is in the electron beam.

### 3. Results

(a) *Na$_{0.75}$CoO$_2$*

Na$_x$CoO$_2$ with x=0.75 shows a superstructure (see Figure 1). The superstructure reflections can be described by an incommensurate vector **q**, which is oriented along [110] with a length of 0.135[110]. In real space, 1/**q**=**Q** has a length of 7.4·*d*(110). Because of twinning, this incommensurate superstructure occurs in most diffraction patterns in three directions (see Figure 1b), which complicates the diffraction patterns considerably. Typically one observes the first and second order superreflections only, the latter being much weaker.

(b) $Na_{0.64}CoO_2$

The $Na_xCoO_2$ compound with x=0.64 shows a superstructure in most areas and crystals. Figure 2 shows a typical diffraction pattern. The superstructure can be described with two incommensurate vectors $q_1$ and $q_2$ of equal length and having an angle of 3.6° with the [11-20] and [-2110] directions, as indicated in Figure 2a. The superstructure reflections are stronger than those of the x=0.75 material, but are weaker than those of the x=0.5 compound. Two types of diffraction patterns (see Figures 2a and 2b) are observed, both with the same incommensurate superreflections, in one, extra commensurate superreflections are observed at the $h+1/2,k+1/2,0$ positions, whereas in the other, these reflections are absent. Irradiation with the electron beam, which heats the sample, (see Figure 2c) results in a reduction in the intensities of these sets of superreflections. The incommensurate superreflections fade quickest. The incommensurate modulation vector $Q$ increases in length during the fading.

(c) $Na_{0.55}CoO_2$

The $Na_xCoO_2$ sample with x=0.55 generally shows no superstructure. A few crystals show the diffraction pattern which a common for the x=0.5 compound. One might anticipate for the x=0.5-like superstructure a $q$ slightly smaller than 0.25·[110] due to a sodium composition slightly larger than x=0.5. However, this was not observed. Also in crystals showing no superstructure, electron beam irradiation could not induce a superstructure. On the other hand, starting with crystal areas showing the x=0.5 superstructure, a gradual change in $q$ was observed similar to that of the pure x=0.5 specimen.

(d) $Na_{0.5}CoO_2$

The $Na_xCoO_2$ compound with x=0.5 shows a strong superstructure in all areas. Figure 3 shows typical [001] diffraction patterns. Fig 3a is taken with only a relatively low exposure to the electron beam. The superstructure reflections can be described by a commensurate vector $q$, oriented along [110] and having a length of 0.25·[110]. Since this is a commensurate superstructure, it can also be described with a three dimensional unit cell, which is orthorhombic with $a_o=2a_p$, $b_o=a_p\sqrt{3}$, $c_o=c_p$. The h00 reflections with $h\neq2n$ are absent in diffraction patterns of tilted crystals oriented such that all reflections except for the h00 reflections are weak. This indicates a $2_1$ symmetry along the c axis. Figure 3b is an example of an electron diffraction pattern observed after exposure to an electron dose of $10^6$ electrons per Å$^2$. The superstructure is obviously altered by the electron beam, resulting in a small change in **q** to $(0.25-\partial) \cdot [110]$, where $\partial$ depends on the degree of irradiation, and varies from 0 to 0.01. The change in the superstructure is accompanied with a loss of Na in the irradiated area.

At about 100 K we observed a more complicated diffraction pattern in a few areas, indicating a tripling along the a and the b axes of the orthorhombic cell. An example is shown in figure 3c. It is possible that this is a structural distortion associated with observed changes in the transport and magnetic properties near 88K (6). Because this tripled unit cell could be the local onset of a phase transformation at lower temperature, we also performed experiments at about 20 K. No tripled or other additional superstructure was observed.

In another sample with x=0.5, we also observed the q=0.24 superstructure immediately, and irradiation by the electron beam did change the modulation length only slightly, but the decay in the intensity of the superreflections was similar to those of the q=0.25 areas.

(e) $Na_{0.35}CoO_2$

The $Na_xCoO_2$ sample with x=0.35 generally shows no superstructure. In those cases where a weak superstructure was observed, the diffraction pattern looked similar to that of the $Na_{0.15}CoO_2$ specimen, shown in Figure 5, except for that the superstructure reflections were much weaker for $Na_{0.35}CoO_2$.

(f) $Na_{0.3}CoO_2$

The $Na_xCoO_2$ compound with x=0.3 generally shows a weak superstructure. The superstructure is shown in Figure 4. The superstructure reflections can be described by an incommensurate vector **q**, oriented along [110] and having a length of $0.31 \cdot [110]$.

(g) $Na_{0.15}CoO_2$

The $Na_xCoO_2$ compound with x=0.15 shows a strong superstructure in all areas. Figure 5 shows a typical diffraction pattern. The superstructure can be described with an a√3, a√3 superstructure as indicated in Figure 5. Also in this case irradiation by the electron beam leads to a change: The intensities of the a√3, a√3 superreflections become gradually weaker and diffuse diffraction spots start to appear at the positions with a doubled unit cell (2a, 2a).

**4. Discussion**

We have observed, based on [001] diffraction patterns, a wide range of superstructures for $Na_xCoO_2$, depending on x. They are summarized in Table II. The intensity of the superreflections depends on the composition. For the x=0.5 compound the superstructure is strong, indicating the presence of a well ordered superlattice structure. In those specimens for which the superreflections are generally weak, there are also areas and crystals with no superreflections present. This indicates only a weak tendency to form superstructures at those compositions. Nevertheless, the superreflections are very sharp and lose intensity with increasing scattering angle in the same way as the main reflections, indicating that the superstructure ordering is long range both in the basal plane and along the c axis. Annealing at 200-400 °C did increase the superstructure ordering observed for some compositions, but, for low x, such annealing leads to a collapse of the lattice due to the chemical metastability of the compounds with low x. Thus it should be pointed out that it may be possible to increase the degree of superstructure ordering for compositions other than x=0.5, which is well ordered under our conditions, if annealing conditions can be found that maintain the stability of the phase at various x values.

The superstructures obviously originate in the Na ion-vacancy ordering. In this respect it should be noted that the Na plane in the basic hexagonal structure has two types of Na sites, both of which are only partly occupied. Both Na sites have a prismatic six-fold coordination of O. One Na atom (Na1) is located directly above the Co and the other (Na2) is located at the other prismatic position, above the center of the Co triangles. The relative occupancy of these two sites seems to depend on the composition, since different occupancies are reported for the compounds with x being 0.74, 0.61 and 0.31 with occupancies for the Na1 and Na2 sites of 0.23 and 0.51, 0.17 and 0.44 and 0.0 and 0.31 respectively (see Table I for references). The reported occupancies suggest a preference for the Na2 site.

The superstructures observed show no immediately obvious relation when compared to each other or with the Na composition. A direct relation might be obscured by the presence of twinning. For this it is useful to consider the twinning in $Na_{0.5}CoO_2$. The x=0.5 compound shows the strongest superreflections, and - provided a small diffracting area is chosen - it is clear that the superstructure can be described with a commensurate vector **q** along the [110] direction in the case of a low electron dose and a related incommensurate vector in the case of a longer exposure to the electron beam. When the diffraction area is more than 100 nm in diameter, in general a superposition of three different diffraction patterns is observed, indicating that the twin domains are still small.

Since twinning is likely to occur, and is even more likely for the weaker superstructures, we have to consider the possibilities of either a superstructure in just one direction with very strong twinning, or a superstructure in two directions, for the other diffraction patterns as well. In the first case, the superstructure will resemble that of $Na_{0.5}CoO_2$ but with a different periodicity. In the second case, the superstructure is of a different type, but might be based on the same principle.

One possible mechanism for all observed superstructures might be a distribution of the Na over the available Na sites such that the Na atoms are spaced as far apart as possible. This would, however, suggest strong superstructure ordering for x =1/3 and x=2/3 with a $a\sqrt{3}$, $a\sqrt{3}$ superstructure, for which only one of the two Na sites is partially occupied. The absence of superstructure reflections for the x =1/3 [001] zones used in our analysis might suggest that no ordering exists for these compositions, but this may also be due to a disordered stacking along the c axis, as is discussed for the x=1/3 model below. Structure refinements of the average structure reported in the literature show a partial occupancy of both Na sites, however. Thus another possible guiding principle, at least for some compositions, could be the dominance of a certain characteristic distribution of Na over the two available sites. This might be governed by the interaction of the Na positions with the distribution of $Co^{3+}$ and $Co^{4+}$ ions. In any model, the stacking of the Na planes along the *c* axis has to be considered. They could either be 'in phase' or 'in antiphase' in neighboring planes. In the latter case the Na ions in adjacent planes are positioned as far away as possible from the first Na plane.

A likely model for the Na plane for the x=0.5 compound is given in Figure 6c. This model is consistent with the observed $2_1$ axis symmetry along the *a* axis. The model is also consistent with the mechanism of maintaining the largest distance possible between the Na atoms, given the availability and observed occupancies of two types of Na sites. One out of four Na1 as well as one out of four Na2 sites are occupied. This distribution of the Na atoms leads to an orthorhombic superstructure with $a=2a_h$, $b=a_h\sqrt{3}$. In order to realize the $2_1$ symmetry, the adjacent plane has to be as indicated in Figure 6c. Along the *a* axis, lines of filled Na1 or Na2 sites parallel the *b* axis are in sequence with lines of vacancies. A change in Na composition can be accommodated easily by increasing the number of adjacent lines of vacancies or lines of Na. For instance, with one Na line and two vacancy lines, x would be 1/3 (see Figure 6d), whereas by a regular repeat of two different widths of the Na lines (see Figure 6e) a much larger Q or an incommensurate occupancy modulation can be obtained. For a given incommensurate modulation being w times as large as the d(110) spacing, the occupancies will be (w-n)/w for single vacancy lines and n/w for single Na lines, where n is an integer. As can be seen from Table II, the observed x and q values can be well fitted for a chosen value of n. Note that in this model, the guiding principle is the presence of ordered lines of filled and occupied Na sites rather than the maximized spacing of the Na atoms. No model for the x=0.64 structure is given because this structure must be more complicated due to the deviation of Q from the [110] direction. Also, the absence of diffraction patterns of this composition with a single modulation direction even for spot sizes as small as about 5 nm, could indicate either a very strong twinning of the Na/vacancy lines along the c axis (rotating by ±120° between adjacent layers), or a real two dimensional modulation in the a-b plane.

The unit cell of the superstructure is determined by the width of the repeat of the lines of vacancies and Na. The 'choice' of occupying the Na1 or Na2 position in the Na lines and the stacking of the ordering in the Na planes will determine the space group and the intensities of the superreflections. The variation in the intensities of the superreflections for various values of x, as well as differences in the systematic absences for one value of x, all indicate that the ordering along the c axis is variable. This variation will be in the relative positions of the Na/vacancy lines on adjacent layers, but also can be that the directions of the Na/vacancy lines are different in adjacent layers.

Models for the stackings along the *c* axes of the various superstructures (except for the x=0.64 specimen) are given in Figure 6. The models are based on the reported occupancies of the Na1 and Na2 sites for some compositions (see Table I), the modulations observed, and a good distribution of the Na atoms in the Na planes.

With x ranging from 0 to 1, the valence of Co varies from 4+ to 3+. It is likely that the distribution of $Co^{3+}/Co^{4+}$ and $Na^+/\square$, where $\square$ is a vacancy, are coupled. The most direct coupling would be by occupying the Na1 sites only and having Na1-$Co^{3+}$ and $\square$1-$Co^{4+}$ strings along the c axis. But the Na1 is the less preferred site.

The sample with x=0.35 shows no superstructure. This could be due to the absence of any ordering in the Na planes, which would be rather remarkable, given the ordering for all other compositions. More likely is that there is a disordering in the stacking of ordered Na planes along the c axis. This model for composition x=1/3 is given in Figure 6e. If the occupied Na sites in neighboring Na planes are shifted at random over 1/3,2/3 or 2/3,1/3 of the superstructure cell, on average, no superstructure will be visible in [001] diffraction patterns. Note that although one can have still 3-fold symmetry in one Na plane, the combination of two adjacent Na planes has no longer 3-fold symmetry. Note also that three equivalent positions for the Na atoms in the adjacent Na plane can be chosen. Continuing this freedom to the next Na plane, the presence of a strong disorder along the c axis is quite logical.

A similar effect can explain the a√3, a√3 superstructure observed for x=0.15. An ordering in the Na planes will create a 3a, 3a supercell for which x=0.11 (see Figure 6f). By introduction of an ordering scheme such that in a given Na plane the Na atoms are in the middle of the projections of three Na atoms of the neighboring plane (requiring a shift of 1/3, 2/3 or 2/3,1/3) and a random 'choice' for either of these two possibilities, the [001] diffraction patterns will show the a√3,a√3 superstructure instead of 3a,3a one.

The uncertainty in the stacking along the *c* axis can in principle be easily solved using electron diffraction or high resolution electron microscopy information from specimens with the *c* axis in the imaging plane ([hk0] zones). Unfortunately all these materials are plate-like, allowing for the examination of very few thin areas with this orientation. In such areas we observed diffraction patterns with diffuse streaks along the c* direction, which could be real, but we think they are due to the fact that these areas are very close to the edge of the specimen. Indeed, nano-diffraction experiments on [001] oriented crystals show the absence of superreflections in the areas close to the edges. Ion milling and cutting with a focused ion beam have the disadvantage that the ion beam can result in changes in the superstructure, an example of which is shown in Figure 7. Ultramicrotomy is planned to obtain [hk0] diffraction information from the interior of the grains.

The superstructure can be changed by the electron beam. EDX elements analysis shows that this change is accompanied by changes in both the Na and the O content. The changes in the superstructures are 1) a gradual loss in intensity of the superstructure reflections, 2) a change in the length of the modulation vector for some compositions and 3) the appearance of extra reflections that are quite diffuse. In all cases it was not possible to change the superstructure by irradiation into one observed for specimens with a lower Na content. In the model proposed above, the increase of the modulation vector upon irradiation requires an increase of the width of the vacancy lines, since an increase of the width of the Na lines is not possible (this would lead to an increase of the Na content). Indeed dynamic diffraction calculations using MacTempas software [13] for the model of $Na_{0.5}CoO_2$ for the model given in Figure 6 and an incommensurate model based on the insertion of an extra Na plane, give a reasonable fit with the observed electron diffraction patterns (see Figure 8). The trend of an increase in Q with irradiation is rather logical for x<0.5, but for x>0.5 one would expect a reduction of the modulation vector due to the reduction of the width of the Na lines.

The x=0.5 samples show an insulating behavior, whereas the other compositions all show a metallic behavior. Compared to the other compositions, the x=0.5 samples have two structural features that are different: a) the superreflections are stronger for the x=0.5 phase and b) the x=0.5 phase has two distinctly different Co sites in the required ratio of 1 to 1, whereas the others (except for x=1/3) have more than two positions and in ratios that do not match the $Co^{3+}/Co^{4+}$ ratio. Both differences are likely coupled and must be the cause of or at least essential to the insulating properties, and that a 'better' ordering for the other compositions might make those compounds insulating as well.

The change in the Na content and the lengths of the modulation vectors upon irradiation by the electron beam indicate a high mobility of the Na atoms. This suggests that, although the sample preparation is different for the various specimens, the mobility of the Na atoms during the removal of the Na by the oxidizing agents is large enough that the formation of well ordered superstructures would occur if it were favored. Thus for compositions other than x=0.5, the driving force to a well ordered Na ordered superstructure must be very small. Possibly thermal annealing of the various compounds can increase the ordering of the x≠0.5 phases, however, both the transport measurements [6] and the present results suggest that the high degree of ordering at x=0.5 may be a consequence of coupling of the Na positions to a charge ordered state of the $Co^{3+}$ and $Co^{4+}$ at a ratio of 1:1 in the $CoO_2$ plane.

**5. Conclusions**

Study of the $Na_xCoO_2$ compounds by electron diffraction has revealed a surprisingly complex variation in Na ion ordering across the phase diagram. The Na ion positions in the ordered structures are not apparently determined solely by maximizing distances between neighboring ions within the plane, but also by likely by interactions with the underlying Co lattice. At our present level of understanding, the Na ion ordering scheme in the layers does not seem to effect the electrical properties in large regions of the phase diagram, except for the composition $Na_{0.5}CoO_2$. However, the ordering may eventually be proven to influence the properties at other composition as well, where the superlattice is well-developed and there are unusual properties. One such composition may be $Na_{0.75}CoO_2$, where we have observed a well developed superlattice and there is magnetic ordering at low temperatures [14-16]. Our electron microsopy study, coupled with previous measurements of the physical properties, indicate that the triangular lattice conductor $Na_xCoO_2$ offers a wide new range of possible structure-property relationships with a complexity potentially equivalent to the square-lattice based perovskites. Our results further suggest that changes in physical properties at a single Na content depending on whether the Na sublattice is ordered or not may eventually be observed, when schemes for ordering the Na at different compositions are developed fully. The possibility of obtaining single crystals for many of the compositions in the $Na_xCoO_2$ family suggests that detailed work will shed much light on the electronic and structural properties of this new type of conducting layered lattice.

**Acknowledgements**

The work at Princeton University was supported by the NSF, grant DMR 0244254, and the DOE, grant DE-FG02-98-ER45706. The work at Delft was supported by the Nederlandse Stichting voor Fundamenteel Onderzoek der Materie (FOM)..

**Table I.** Reported basic structures for $Na_xCoO_2$, including the occupancies of the Na ions. The Na 1 site is located directly above the Co atoms in their triangular layers. The Na2 site is located in above the center of the triangle formed by three Co atoms. The symbol / in the stacking column indicates the location of the Na atom planes. A, B, and C refer to the different types of close packed oxygen layers.

| $x$ | Stacking | Cell parameters in nm | Space group | O(Na1) | O(Na2) | Ref. |
|---|---|---|---|---|---|---|
| 0.74 | A/AB/B | a=0.284, c=1.081 | $P6_3/mmc$ | 0.23 | 0.51 | [11] |
| 0.67 | A/BC/AB/C | a=0.490, b=0.283, c=0.572, ß =105.96° | C2/m | - | - | [10] |
| 0.61 | A/AB/B |  | $P6_3/mmc$ | 0.17 | 0.44 | [12] |
| 0.60 | A/AB/BC/C | a=0.283, b=0.484, c=1.653 | R3m | - | - | [8] |
| 0.31 | A/AB/B |  | $P6_3/mmc$ | 0 | 0.31 | [12] |

**Table II.** Observed superstructures and the changes in the superstructures upon electron beam irradiation for various compounds $Na_xCoO_2$. Also the expected composition based on the modulation model described in the text is given.

| x | Observed superstructures Modulation (Q) length (nm) / direction | times d(110) | n for closest comp. | x for this n | Effect of electron beam |
|---|---|---|---|---|---|
| 0.7 | 1.04 [110] | 7.4 | -2 | 0.73 | no change in Q, fading |
| 0.64 | 0.75-0.8   2-4° from [110] | ~5.5 | -2 | 0.64 | increase in Q, fading |
| 0.55 | most crystals no superstructure, some crystals x=0.5 type superstructure | | | | not checked |
| 0.5 | 0.563 [110] | 4 | 2 | 0.50. | increase in Q (->0.59), fading |
| 0.35 | almost no superstructure | | | | |
| 0.3 | 0.471 [110] | 3.3. | 1 | 0.30. | no change in Q, fading |
| 0.15 | 0.423 (a√3,a√3) [110] | | - | - | Q unchanged, fading , extra 2a,2a ss |

Figure 1.

[001] diffraction patterns of $Na_{0.7}CoO_2$ showing the superstructure of a) an almost single crystalline area, and b) a twinned area. The basic structure reflections 100 and 010 are indicated. White arrows indicate the first and second order superreflections of the central spot.

Figure 2.

The two types of [001] diffraction patterns observed for $Na_{0.64}CoO_2$. Both show incommensurate superreflections but they differ in a) showing and b) not showing the presence of reflections at 1/2,0,0 and similar positions. Some of these reflections are encircled in b). The incommensurate vectors $q_1$ and $q_2$ of equal length and having an angle of 3.6° with the [11-20] and [-2110] directions are indicated with white arrows in a). c) shows a diffraction pattern after irradiation by the electron beam.

Figure 3.

[001] diffraction patterns of $Na_{0.5}CoO_2$. a) was taken almost immediately, and shows a commensurate superstructure, b) was taken after exposure to the electron beam, leading to an incommensurate superstructure. c) shows a diffraction pattern taken at about 100 K with extra reflections; such diffraction patterns were only observed in a few areas. White arrows indicate the first and second order superreflections of the central spot and the 110 spot.

Figure 4.

[001] diffraction pattern of $Na_{0.30}CoO_2$ showing a superstructure with for x=0.3. White arrows indicate the first and second order superreflections of the central spot and the 110 spot.

Figure 5

[001] diffraction patterns of $Na_{0.15}CoO_2$ showing an ($a\sqrt{3}$, $a\sqrt{3}$) superstructure. The average structure 100 type reflections are circled. a) was taken shortly after irradiation by the electron beam. b) was taken after a considerable electron beam irradiation. Faint reflections at positions half way between neighboring average structure reflections can be observed.

Figure 6.

Models for the Na planes of $Na_xCoO_2$ with x being 0.75, 0.71, 0.5, 0.33, 0.30 and 0.15. Both the Na planes on z=0 and z=1/2 are given. Occupied Na positions are given as black dots and Na vacancies by open circles. X indicates the positions of the underlying Co atoms. On the right of each model it is indicated whether the Na atoms are at a Na1 or a Na2 site. The rows of Na atoms perpendicular to the modulation direction are indicated with arrows under each model and the modulation frequency is given above each model. The basic hexagonal unit cell is indicated by dotted lines, the observed unit cell by full lines and possible repeat in individual Na planes by dashed lines.

Figure 7

[110] electron diffraction pattern of a thin specimen of $Na_{0.75}CoO_2$ prepared by cutting with a focused ion beam. A large number of extra reflections are present, which are probably artifacts due to modifications by the ion beam.

Figure 8

Calculated diffraction patterns for specimen thicknesses of 5 nm and for the models for the Na positions given above the diffraction patterns. a) and b) show the diffraction patterns for the commensurate and incommensurate modulations respectively. The vacancy lines are indicated in the models. For the incommensurate modulation, one extra vacancy line is present, associated also with a shift over 1/2b of the Na lattice, as indicated by the arrow.

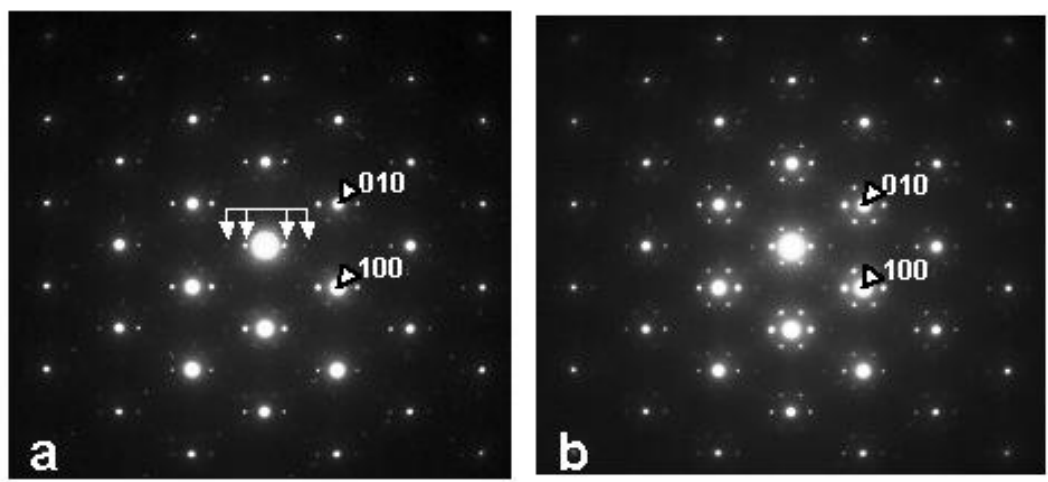

Figure 1.

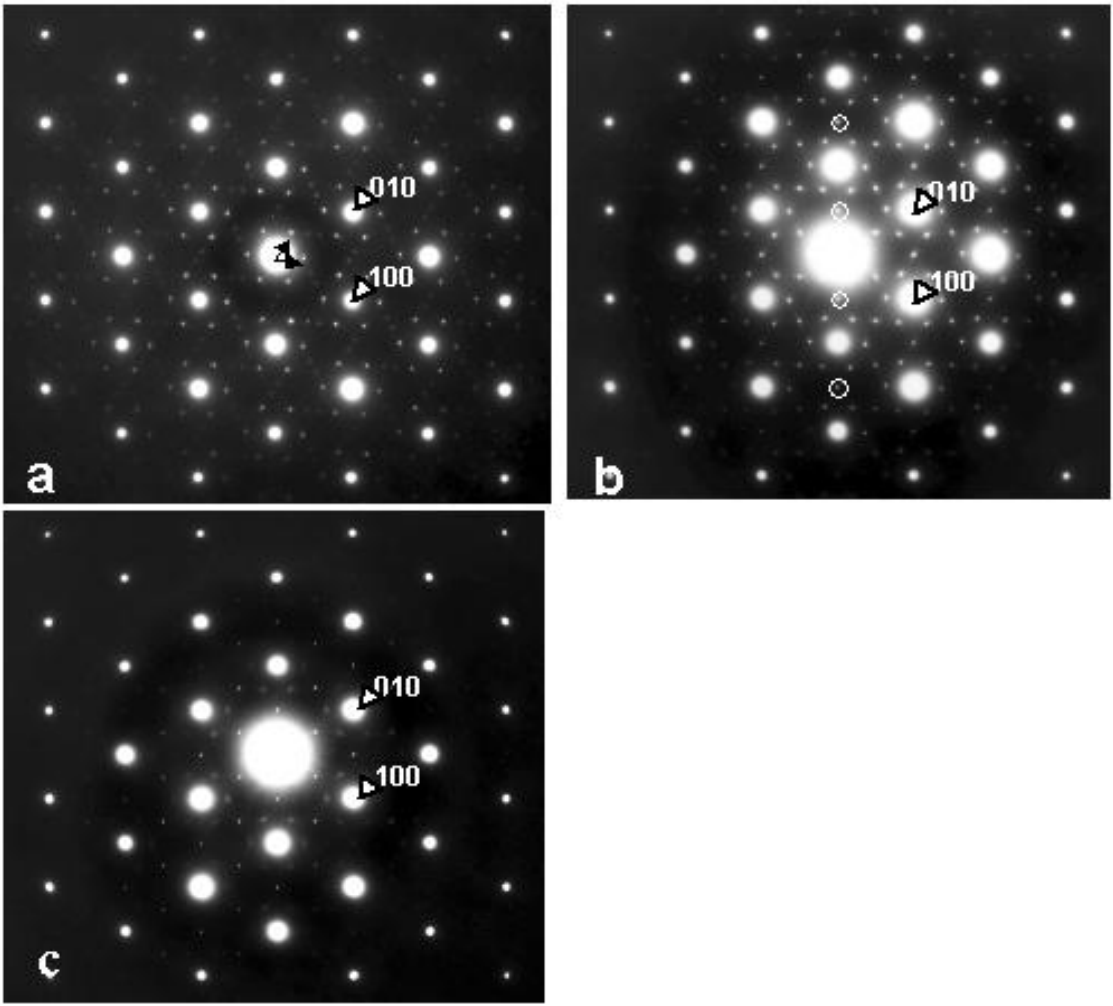

Figure 2.

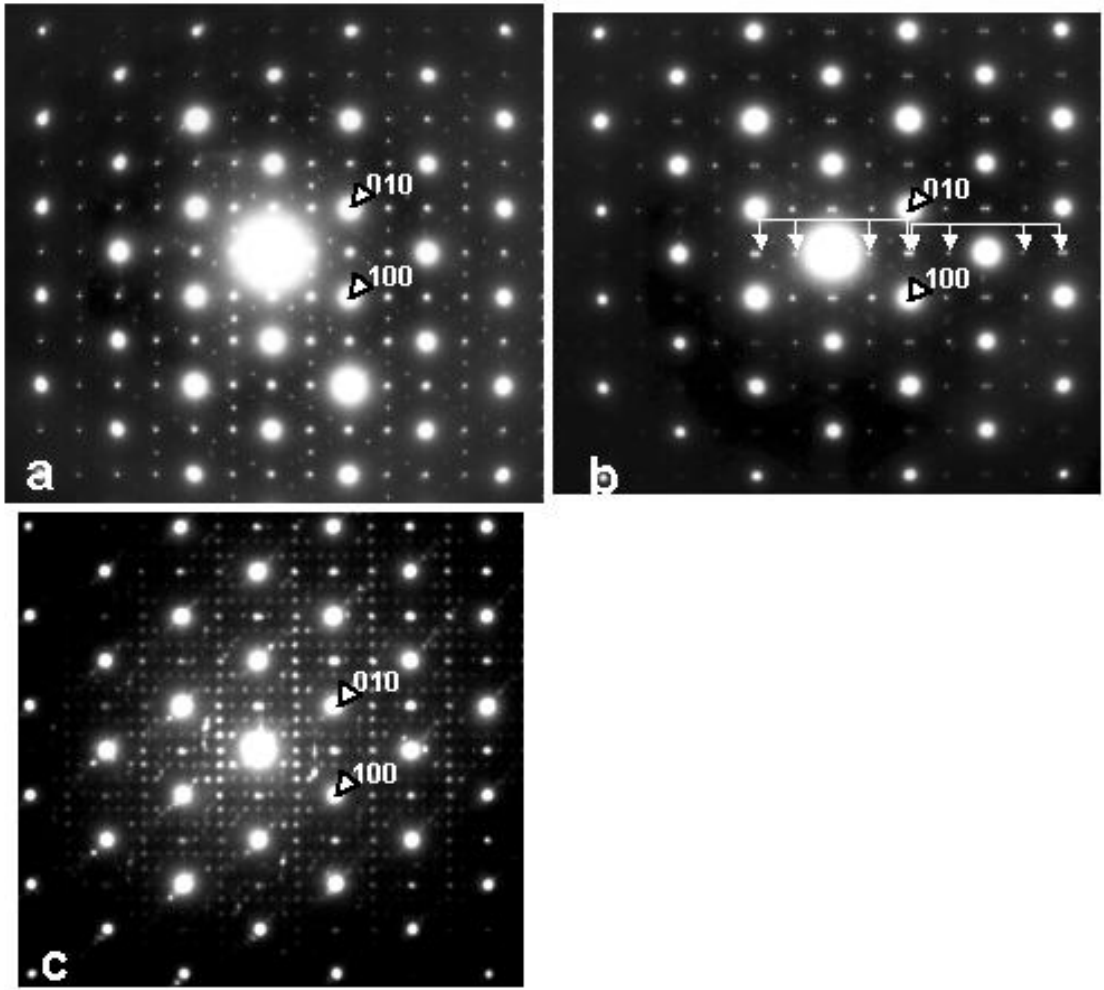

Figure 3.

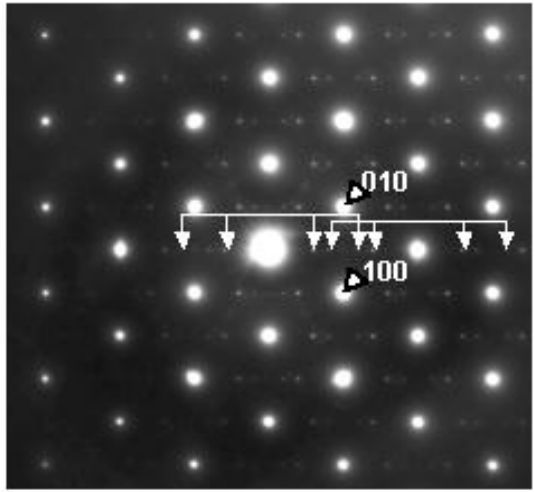

Figure 4.

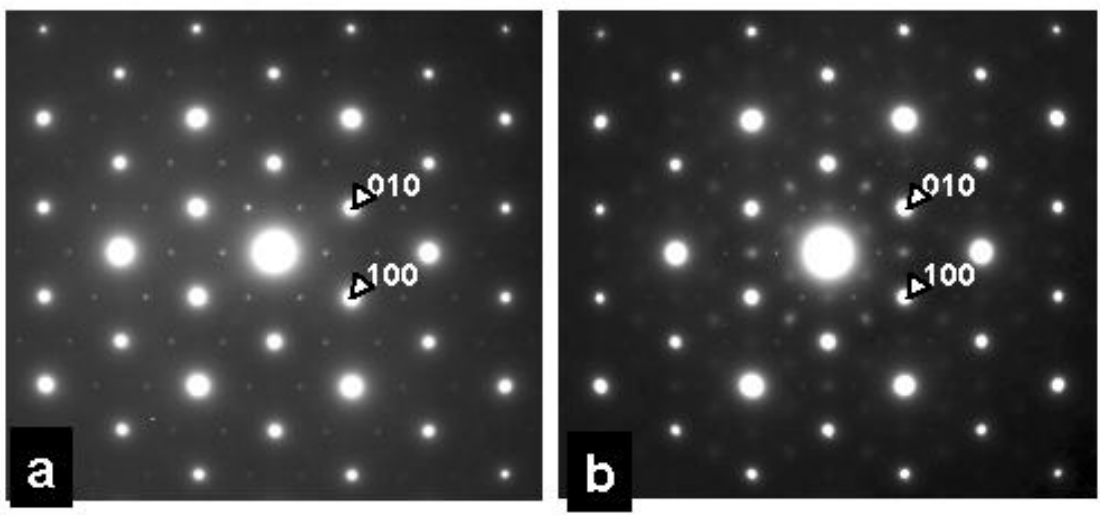

Figure 5.

## Model x=.75

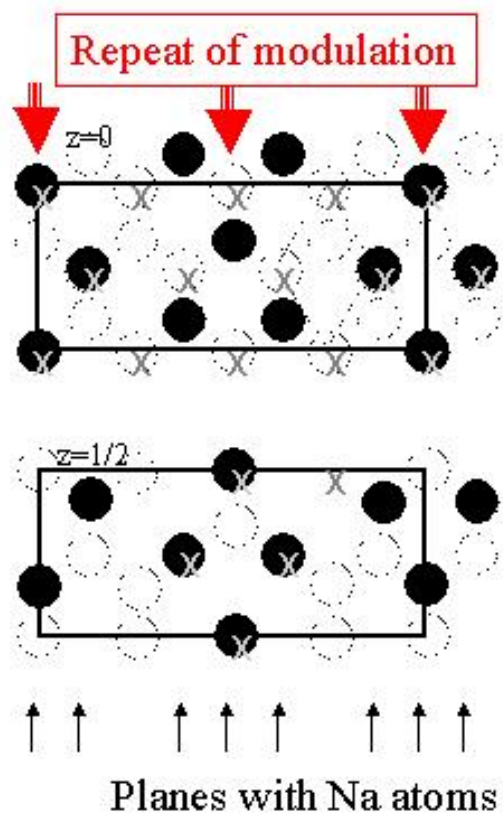

Figure 6(a)

Planes with Na atoms

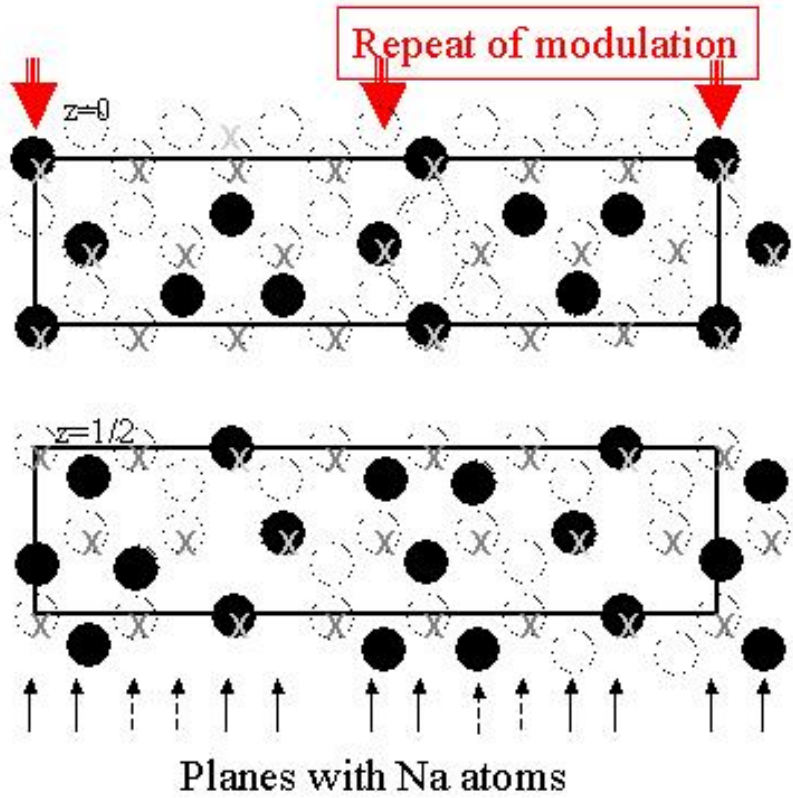

Figure 6(b)

## Model x=.5

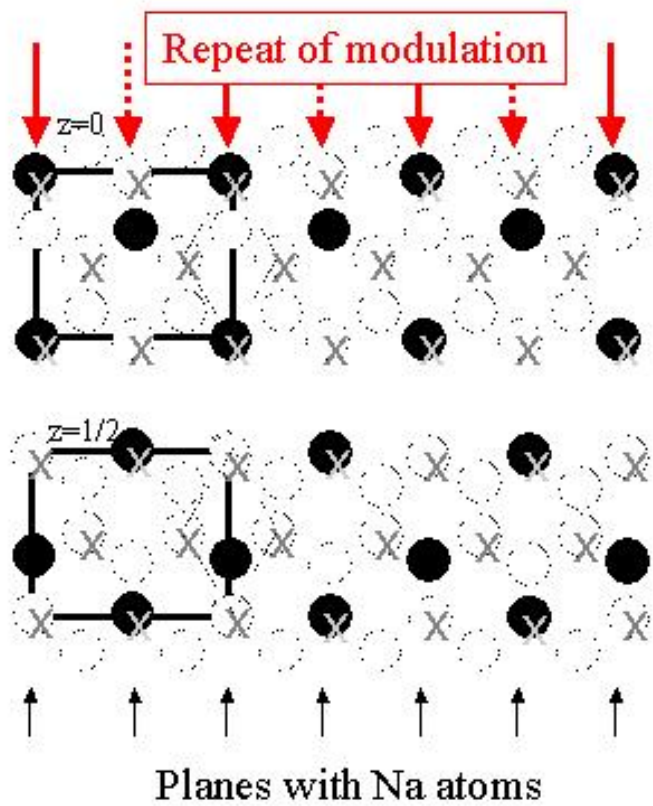

Figure 6(c)

# Model x=.33

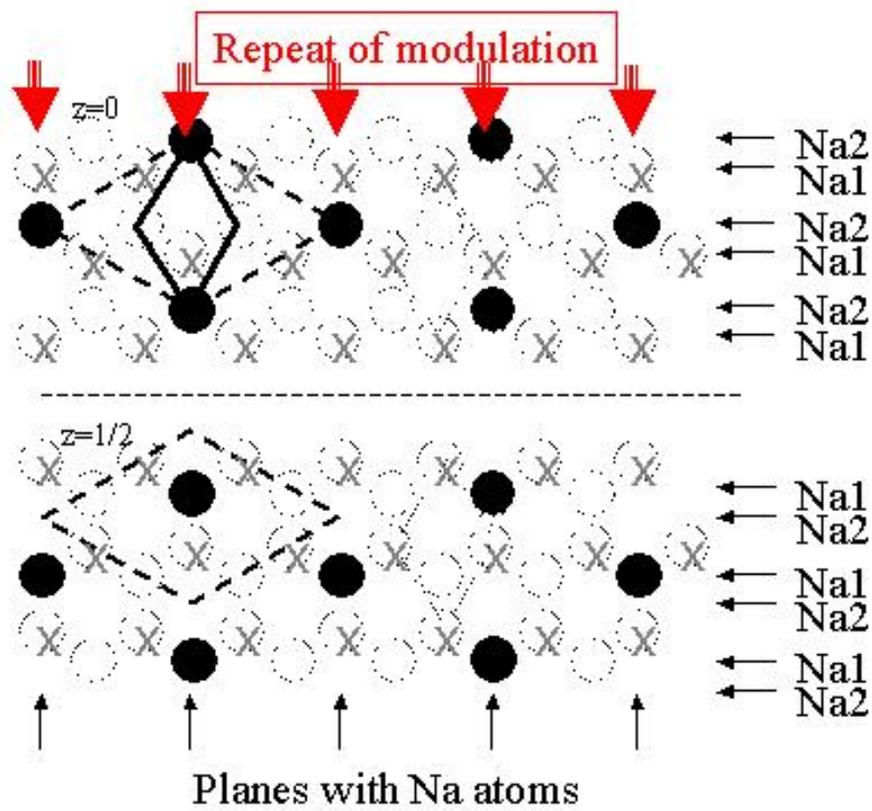

Figure 6(d)

# Model x=.3

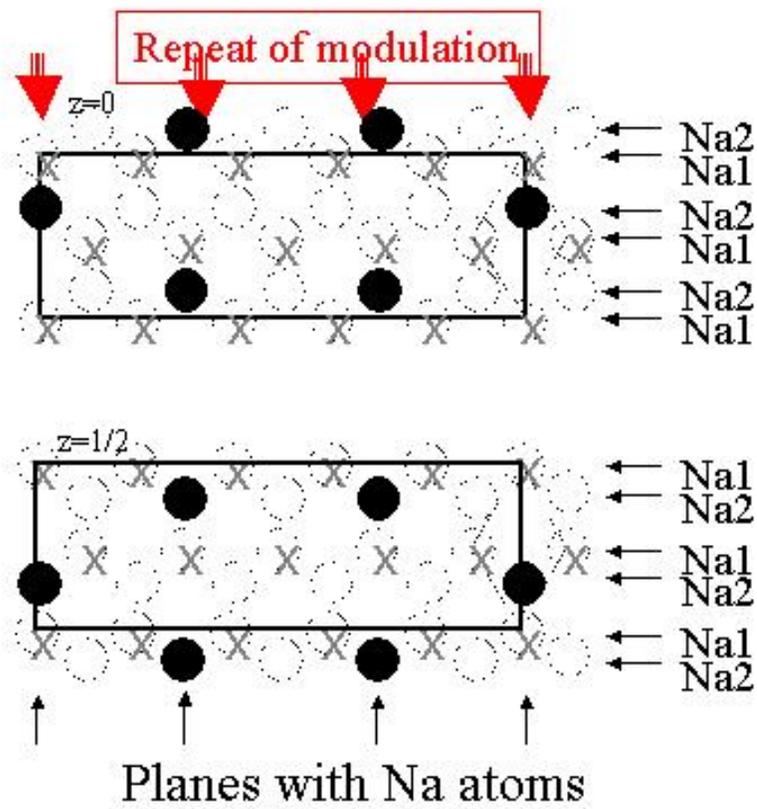

Figure 6(e)

Planes with Na atoms

## Model x=.11

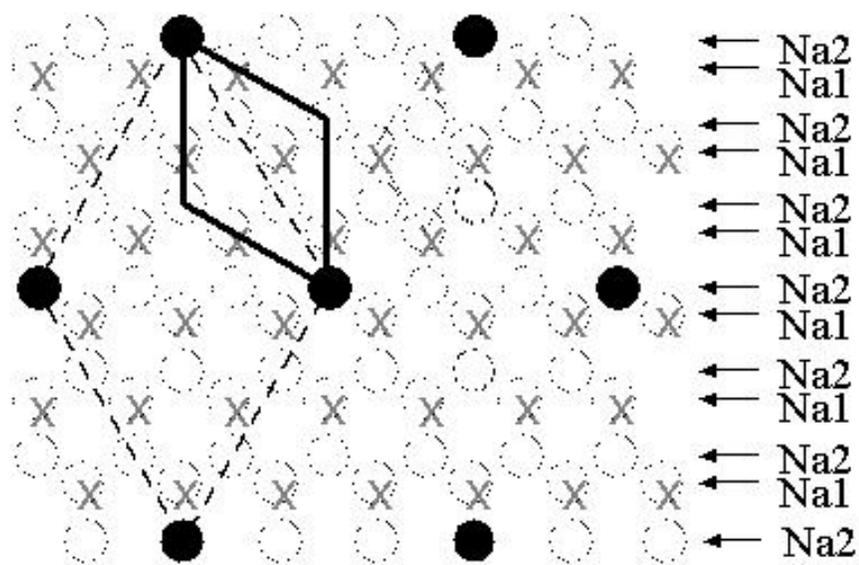

Figure 6(f)

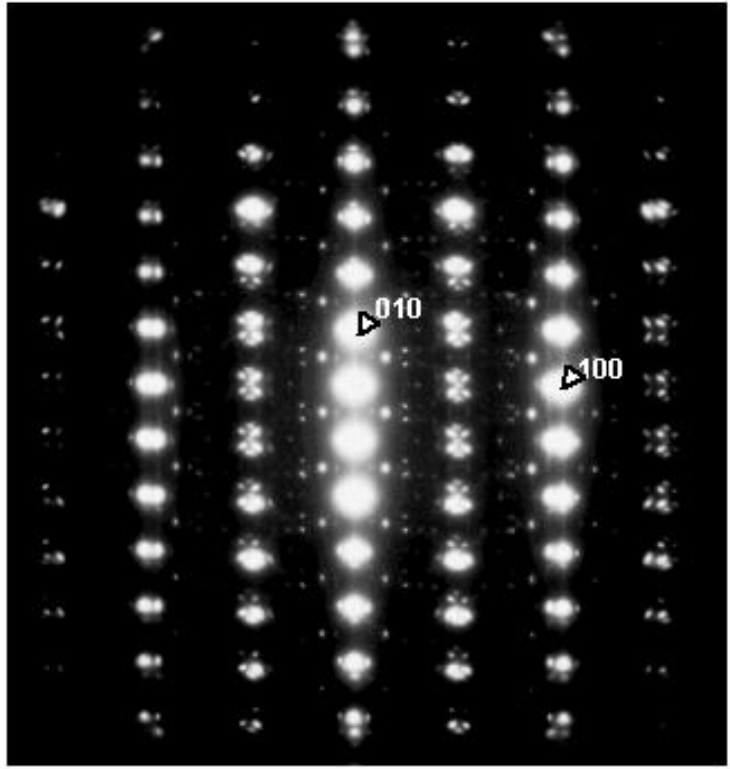

Figure 7.

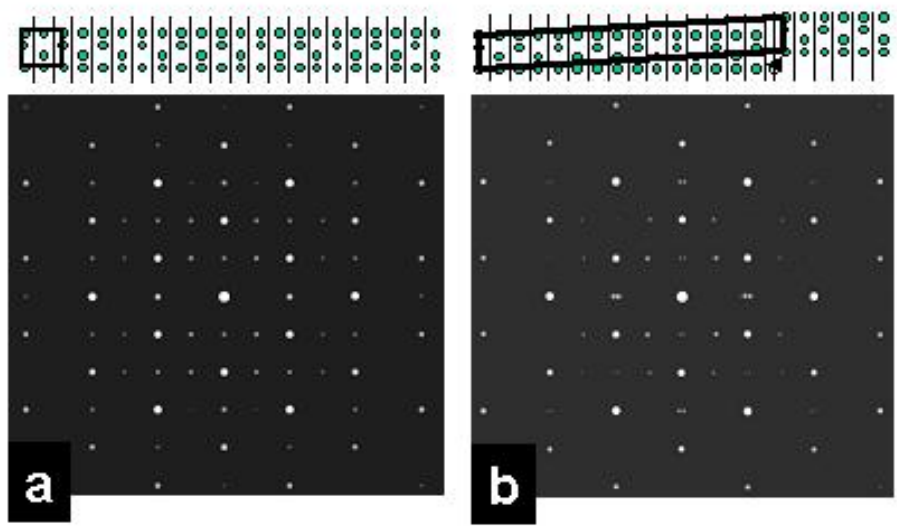

Figure 8.